\documentclass[pdflatex, sn-nature, referee]{sn-jnl}

\usepackage{graphicx}
\usepackage{geometry}
\usepackage{multirow}
\usepackage{amsmath,amssymb,amsfonts}
\usepackage{amsthm}
\usepackage{mathrsfs}
\usepackage[title]{appendix}
\usepackage{xcolor}
\usepackage{color}
\usepackage{soul}
\usepackage{textcomp}
\usepackage{manyfoot}
\usepackage{booktabs}
\usepackage{algorithm}
\usepackage{algorithmicx}
\usepackage{algpseudocode}
\usepackage{listings}
\usepackage{upgreek}
\usepackage{hyperref}

\usepackage{tikz}
\usepackage{xr}
\makeatletter
\newcommand*{\addFileDependency}[1]{
\typeout{(#1)}
\@addtofilelist{#1}

\IfFileExists{#1}{}{\typeout{No file #1.}}
}\makeatother

\newcommand*{\myexternaldocument}[1]{
\externaldocument{#1}
\addFileDependency{#1.tex}
\addFileDependency{#1.aux}
}

\myexternaldocument{SI}

\begin{document}

\title{\large \textbf{Pulsed magnetic field gradient on a tip for nanoscale imaging of spins}}

\author[1]{\fnm{Leora} \sur{Schein-Lubomirsky}}
\author[2]{\fnm{Yarden} \sur{Mazor}}
\author[3]{\fnm{Rainer} \sur{Stöhr}}
\author[3]{\fnm{Andrej} \sur{Denisenko}}
\author*[1]{\fnm{Amit} \sur{Finkler}}\email{amit.finkler@weizmann.ac.il}

\affil[1]{\orgdiv{Department of Chemical and Biological Physics}, \orgname{Weizmann Institute of Science}, \city{Rehovot} \postcode{7610001}, \country{Israel}}
\affil[2]{\orgdiv{School of Electrical Engineering}, \orgname{Tel Aviv University}, \city{Tel Aviv-Yafo} \postcode{6997801}, \country{Israel}}
\affil[3]{\orgdiv{3.\,Physikalisches Institut}, \orgname{Universität Stuttgart}, \city{Stuttgart} \postcode{70569}, \country{Germany}}

\abstract{Nanoscale magnetic resonance imaging (nanoMRI) is crucial for advancing molecular-level structural analysis, yet existing techniques relying on permanent magnets face limitations in controllability and resolution. This study addresses the gap by introducing a switchable magnetic field gradient on a scanning tip, enabling localized, high-gradient magnetic fields at the nanoscale. Here, we demonstrate a device combining a metal microwire on a quartz tip with a nitrogen-vacancy (NV) center in diamond, achieving gradients up to 1 $\upmu\text{T nm}^{-1}$ at fields below 200 $\upmu$T. This allows electron spin mapping with 1 nm resolution, overcoming challenges like emitter contrast and sample preparation rigidity. The current-controlled gradient, switchable in 600\,ns, enhances precision and flexibility. Additionally, the metallic tip modifies Rabi power spatially, enabling selective spin manipulation with varying microwave effects. This innovation paves the way for advanced nanoMRI applications, including high-resolution imaging and targeted spin control in quantum sensing and molecular studies.}

\maketitle

\section{Introduction}\label{sec1}
Nanoscale magnetic resonance imaging (nanoMRI) aims to give insight to molecular structure at the single molecule level, and has advanced significantly over the past decades \cite{Budakian2023}. It has the potential to gain deeper understanding in the vast worlds of chemistry and biology where conventional nuclear magnetic resonance (NMR) is limited by sample size \cite{suter1992sensitivity} and conventional MRI is limited by magnetic field gradients to the micron scale \cite{Ciobanu2002}. The basic principle of MRI is the encoding of spatial information via spectral variation of the spin's Larmor precession frequency. In this way, a known gradient magnetic field is applied over the detection region. This separates the resonance frequency of similar species in the sample via $f(\mathbf{r})=\gamma_i B(\mathbf{r})$, where $\gamma_i$ is the gyromagnetic ratio of the detected species and $B(\mathbf{r})$ is the local magnetic field magnitude. A variety of techniques exist in the field of nanoMRI, among them - magnetic resonance force microscopy (MRFM), electron spin resonance using scanning tunneling microscopy (ESR-STM) and the nitrogen-vacancy (NV) center in diamond. Modern MRFM cantilevers allow detection of on the order of $100$ molecules \cite{Budakian2023} with an impressive 10 nm resolution \cite{Nichol2013, Degen2009}. Moreover, and specifically relevant to our work, a recent method of applying the gradient was implemented by focusing the current in a planar device below the cantilever, resulting in resolution below 2 nm \cite{Rose2018}. ESR-STM is a surface technique and is limited by the type of systems that can currently be explored \cite{Baumann2015}. Together with ESR-AFM \cite{Sellies_2023}, MRFM and ESR-STM are all technically very challenging.

\begin{figure*}[!ht]
    \centering
    \includegraphics[width=0.9\textwidth]{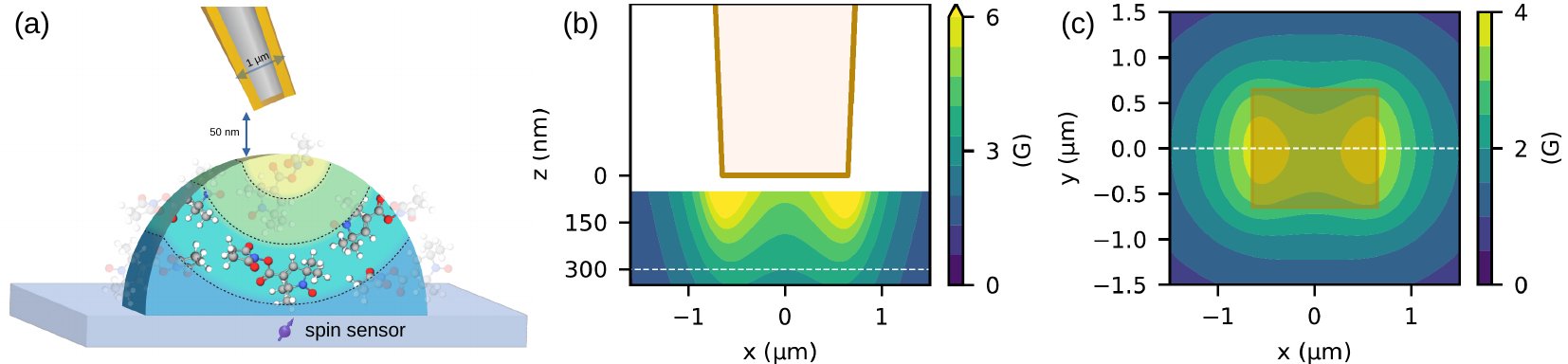}
    \caption{\textbf{Scheme and simulation of magnetic gradient on a tip}. (a) Schematic view (not to-scale) of the magnetic gradient induced by the focused current in the vicinity of the tip. (b, c) Simulation of the magnetic field for a tip with an apex of 1300 nm diameter. (b) In-plane cross-section of the magnetic field below the tip at the center of the tip. The tip is shown in the figure ending at $z=0$. The white dashed line is the line-cut shown in panel c. (c) Cross-section simulation of magnetic field at a distance of 300 nm from the tip apex. The brown rectangle shows the cross-section of the tip above the surface. The white dashed line is the line cut shown in panel b.}
    \label{fig: schematic}
\end{figure*}

In NV magnetic resonance sensing, the spin state of the NV interacts with magnetic fields in its immediate environment and the field is measured via optical detection of the NV spin state \cite{Gruber1997}. Nanoscale resolution can be achieved by sensing single NV defects which can detect local magnetic fields and have demonstrated AC sensitivity better than 10 nT Hz$^{-1/2}$ \cite{Balasubramanian2009}, comparable to sensing the magnetic field of a single electron at a distance of 50 nm \cite{Grinolds2013}. Further experiments have detected NMR signals with nanoscale sample size detecting as few as $10^{4}$ nuclear spins \cite{Staudacher2013,Mamin2013}, and later on reaching the single nuclear spin limit \cite{Sushkov2014, Mueller2014}. In these experiments, the signal is integrated over all spins interacting with the NV, and one is unable to differentiate between individual spins. One can use the gradient created by the NV center's own dipole moment, thereby modifying the Larmor precession frequency of nearby spins due to the hyperfine interaction \cite{Taminiau2012, Kolkowitz2012, Zopes2018}. However, switching the gradient on and off in this case depends on the NV's spin state and in addition the gradient itself cannot be moved around. Another gradient-based method with high spatial resolution, similar to MRI, was demonstrated by Grinolds et al.\,\cite{Grinolds2014}, where sub-nm resolution was achieved by depositing a thin ferromagnetic film on a tip which applied a local gradient magnetic field \cite{Balasubramanian2008}. The spatial gradient gives rise to spectral separation that encodes the location of nearby molecules within the NV sensing region. This impressive milestone is limited by the strong off-axis magnetic fields induced by the ferromagnet, as such fields are detrimental to NV contrast due to mixing of the zero-field splitting (ZFS) eigenstates. The ZFS defines the NV's $z$-axis  \cite{Tetienne2012}. This can be seen in the ground-state spin Hamiltonian of the NV center,
\begin{equation}
    \label{eq: Hamiltonian}
    \mathcal{H} = \hbar D_\mathrm{zz}S^{2}_z+g\mu_B\left(B_z S_z + B_\perp S_\perp\right),
\end{equation}
with $D_\mathrm{zz}$ being the ground-state ZFS, $B_z$ the magnetic field along the N$\rightarrow$V axis, $\mu_B$ is the Bohr magneton and $g$ is the Land\'e g-factor for the NV. Specifically, it is these off-axis ($B_\perp$) fields that are created by a ferromagnet that are detrimental to the initialization and readout stages of the NV's pulse sequence used for sensing.

Current-based pulsed magnetic fields can overcome this challenge, as they can be switched on and off throughout the experiment. Thus, the magnetic field is applied during the sensing stage of the pulse sequence but switched off during spin read-out, which is optimal when the NV eigenstates are prominently defined by ZFS and $B_z$ terms. Switching the current using an electromagnet was demonstrated by Bodenstedt et al.\,\cite{Bodenstedt2018} where a hard drive's write head was used to achieve magnetic gradients up to 100\,$\upmu\text{T nm}^{-1}$, yet also here magnetic noise from the write head's electromagnet limited the detection contrast. An additional advantage of pulsed magnetic fields is that they allow one to measure small shifts that may be masked by drifts in the system. With a switchable field, one can sequentially measure a signal with and without the magnetic field, thus becoming indifferent to system drifts on longer time scales. Lastly, this allows unique pulse sequences that leverage a changing gradient within an experiment, such as Fourier imaging \cite{Arai2015,Amawi2023,Wang2023} and gradient pulse sequences \cite{Stejskal1965, Shemesh2013}. Therefore, a switchable magnetic field gradient is a requisite for atomic-scale NV-MRI. 

In this work, we demonstrate a device tailored for nanoMRI, that provides an on-demand localized magnetic field gradient on the apex of a quartz tip (see Fig.\,\ref{fig: schematic}a), reaching gradients as high as $1\,\upmu\text{T nm}^{-1}$, as measured by a nearby single NV center. We demonstrate how this tip, deposited with gold, modulates the effective Rabi driving power of the NV center by a factor of $\times 3.5$, effectively serving as a Rabi power gradient.

\section{Results}
\subsection*{Current focusing device design}\label{sec2}
Inspired by the MRFM community \cite{Rose2018}, the device we present was developed in order to achieve controllable magnetic field gradients localized around a single NV, which acts as a magnetic sensor. To this end, we fabricate a wire by physical vapor deposition (see \hyperref[sec11]{Methods}) that runs along a quartz tip. The abrupt change in the wire's angle at the apex of the tip results in a gradient in the magnetic field induced by the current running through the wire. The field of such a configuration is estimated by numerical integration according to the Biot-Savart law (see Supplementary Note 1).

\begin{figure*}[!ht]
    \centering
    \includegraphics[width=\textwidth]{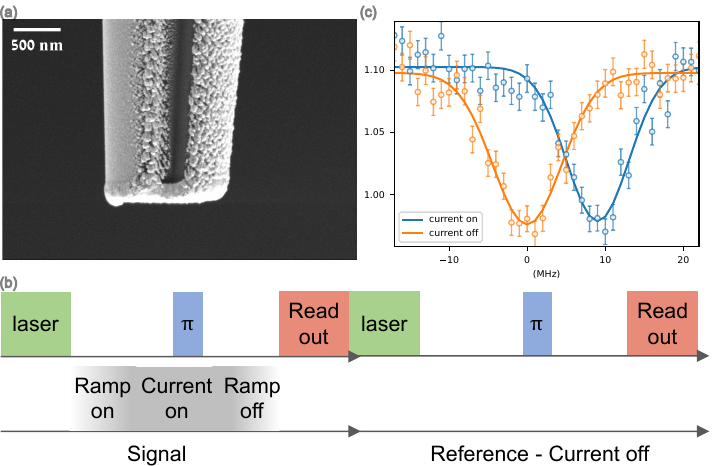}
    \caption{\textbf{Device and pulse sequence}. (a) SEM image of the tip. The quartz substrate (dark region in the center) is coated with gold in a three-layer self-aligned deposition (see Fig.\,\ref{fig: methods process} and \hyperref[sec11]{Methods} for more details). (b) A schematic of the pulsed tip ODMR sequence used to characterize the gradient. A standard ODMR pulse sequence is interleaved with a current pulse (1.54\,mA) through the tip in the first step, and an additional reference step is repeated without the current in the tip. (c) A single (AFM) pixel ODMR measurement, where a measurement is done with and without a current applied to the tip.The error bars are calculated as shown in Supplementary note 4. }
    \label{fig: setup}
\end{figure*}

The simulation results are presented in Fig.\,\ref{fig: schematic}(b,c), showing the magnitude of the magnetic field in-plane and out-of-plane relative to the tip. The field strength is weak relative to the zero field splitting, and so this field is not expected to impact the eigenstates of the NV \cite{Tetienne2012}. We considered a working point of 300\,nm as a stand-off distance between the tip and the diamond, where the maximum field strength is 0.37\,mT (see Supplementary Note 3). The maximum gradient will determine the spatial resolution, calculated to be $2.78\,\upmu\text{T}\,\text{nm}^{-1}$ when running a current of 2\,mA. The final resolution of the system depends also on the spectral resolution, or linewidth, $\Delta\nu$ of the sensor, for a typical single NV setup at ambient conditions this is around 100 kHz for shallow implantation, corresponding to a dephasing time of $T_2^*$ of $\sim 3\,\upmu\text{s}$, where $\Delta\nu = 1/(\pi T_2^*$) \cite{Edmonds2021}. Thus, in order to achieve 1\,nm spatial resolution for electron spins (with the electron's gyromagnetic ratio taken to be $\sim 28\,\text{GHz T}^{-1}$) a gradient of $3.57\,\upmu\text{T}\,\text{nm}^{-1}$ is required, similar to the order of magnitude of the calculated gradient. 

We found the maximal current in the device is limited due to Joule heating induced by the current, which can cause device failure, i.e., open circuit. In order to estimate the heating of the device, we simulate the heating using COMSOL to solve the heat transfer equations for the device geometry (see Supplementary Note 7). From the perspective of heating, a larger cross-section is beneficial to sustain increasing currents (see Fig.\,\ref{fig:temperature}), thus a larger apex will prevent device failure due to heating. Nevertheless, our simulations (as shown in Fig.\,\ref{fig: schematic}) suggest that a smaller apex diameter leads to a stronger gradient. The chosen geometry, therefore, of the device, was set to balance the need for low heating while still achieving a gradient that is beneficial for magnetic resonance imaging. To this end, the tip is designed to have an apex on the order of 1\,$\upmu$m and a deposition thickness on the order of 100\,nm (see \hyperref[sec11]{Methods}).

\subsection*{The gradient sensing setup}

For magnetic resonance sensing, the tip and NV must be co-aligned such that the magnetic field of the tip is positioned optimally relative to the specific NV used as a sensor. Single NVs are embedded in a diamond membrane (30\,$\upmu\mathrm{m}$ thick) with pillars etched on the surface for improved photon collection. With this tip-diamond geometry, a sample can be placed directly on the diamond membrane, such that the tip is oriented above the sample. The diamond sits on a microwave (MW) co-planar waveguide necessary for pulses that manipulate the spin state. An SEM image of the tip with the current-focusing device is shown in Fig.\,\ref{fig: setup}a. It consists of two leads connected at the apex, such that the current flows across the apex (fabrication is described in \hyperref[sec11]{Methods}). The tip is placed on a piezoelectric stage; stick-slip motion is used to control the lateral position of the tip to within 5\,$\upmu\mathrm{m}$ from the NV pillar, and fine motion is used to scan the tip in the vicinity of the NV. To control the vertical distance between the tip and the single NV, the tip is attached to a tuning fork to provide feedback for atomic force microscope (AFM). During the sensing sequence, the tip's vertical position is fixed at a set distance from the diamond surface. This is important to prevent noise in the magnetic field due to the feedback motion of the tip. Alternatively, it can be also moved in-sync with the oscillations of the tuning-fork \cite{Hong2012}.

\subsection*{Magnetic field characterization}
\begin{figure*}[!ht]
    \centering
    \includegraphics[width=\textwidth]{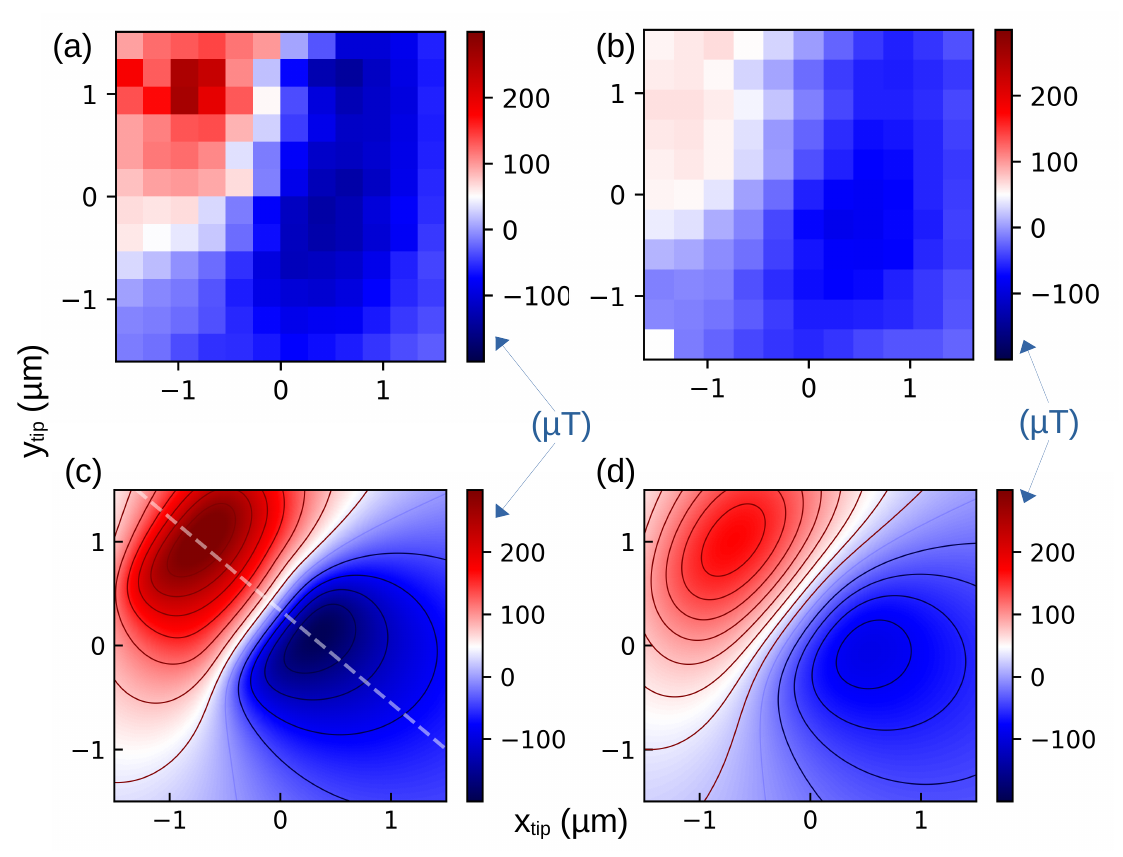}
    \caption{\textbf{Magnetic field generated by the tip}. (a-b) A map of the magnetic field (in $\upmu\text{T}$) of the tip measured by the pulsed tip ODMR sequence from Fig.\,\ref{fig: setup}c at varying distances (300\,nm, 600\,nm) from the diamond surface. (c-d) Numerical calculation of the expected magnetic field (in $\upmu\text{T}$) projected on the NV $[\bar{1}11]$ axis for the conditions measured in (a-b), respectively. The white dashed line in (c) goes along a cross-section of the maximal gradient.}
    \label{fig:magnetic field}
\end{figure*}

To map the DC magnetic field of the tip, an optically detected magnetic resonance (ODMR) pulse sequence, sensitive to magnetic fields that are projected along the NV zero-field splitting axis, is used, where current is applied during the spin state manipulation but turned off during the spin readout and initialization phase, see Fig.\,\ref{fig: setup}c. We identify a rise/fall time of 600\,ns (see Supplementary Note 8 for the MW/tip circuitry). The pulse sequence is alternated with and without current to cancel the effect of slow drifts that would otherwise accumulate and mask small changes in the magnetic field. The tip is scanned relative to the NV and ODMR is measured at each tip location. To avoid damage to the tip, the measurements were performed with a current of 1.54\,mA, as measured on the oscilloscope (see Supplementary Note 5 for a discussion on the current amplitude).

\begin{figure*}[!ht]
    \centering
    \includegraphics[width = \textwidth]{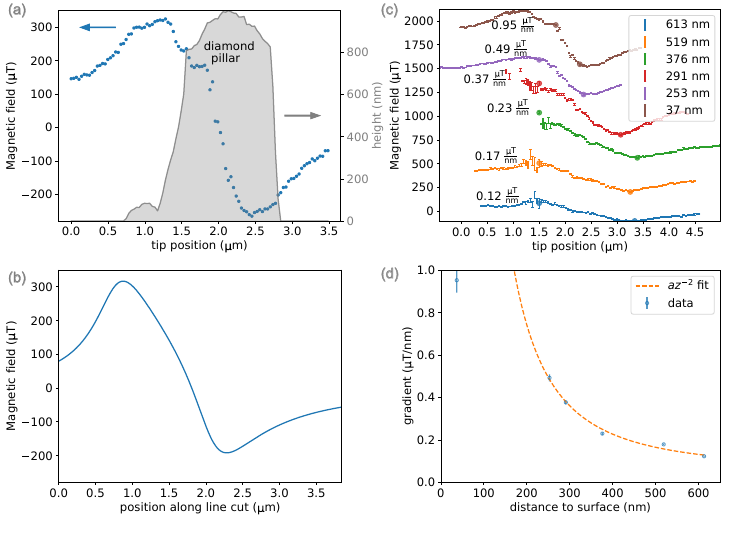}
    \caption{\textbf{Magnetic field gradient along the NV axis}. (a) A line scan across the magnetic field map over the region with the highest expected gradient (0.95\,$\upmu\text{T nm}^{-1}$) from (c). The gray region shows the diamond nanopillar topography measurement to show the location of the steepest slope with respect to the nanopillar position. (b) A plot of the line-cut from the simulation in Fig.\,\ref{fig:magnetic field}(c), in agreement with (a). (c) Line scans measured at varying distances from the diamond surface, the values and errors are extracted from a Gaussian fit to each ODMR. The values for the magnetic field are shifted for clarity. For each line scan, the gradient is calculated as a linear fit along the data between the two circular points. (d) The extracted gradient from (c) as a function of the distance from the surface alongside an inverse quadratic fit. The error is obtained from the error in the gradient fit.}
    \label{fig:line scan}
\end{figure*}

The magnetic field measured at a single pixel (tip position) is shown in Fig.\,\ref{fig:magnetic field}, with each resonant dip fitted to a Gaussian. The shift in the magnetic field due to the tip is calculated as the difference between the center dip measured with and without the current. The magnetic field is then measured over a lateral range of $3\,\upmu\text{m}$ around the pillar with an NV. Scans performed at varying distances to the diamond's surface are shown in Fig.\,\ref{fig:magnetic field}(a,b). The maximum measured field is 259\,$\upmu\text{T}$, which does not hinder the NV contrast \cite{Tetienne2012}. To compare to the simulation of the expected field (as shown in Fig.\,\ref{fig: schematic}c), we account for the degrees of freedom in the measurement. The first is the relative angle between the NV axis to the direction of the current, and the other being the lab coordinates along which the tip is scanned. The magnitude and range of the measured field are affected by the angle between the current to the NV, since the ODMR measurement is most sensitive to fields parallel to the NV axis. While the scanning axis defines the axis connecting the maxima and minima, it does not affect the field strength. The simulation, as shown in Fig.\,\ref{fig:magnetic field}, projected along the NV axis $[\bar{1} 1 1]$ and rotated by 45$^{\circ}$, is qualitatively similar to the measurement. Several features are visible in both - a sharp peak in the field along with a broad yet shallow dip. As expected, the field strength increases with the distance to the surface. Quantitatively, the range and values of the field in the experiment and the simulation are in good agreement. The discrepancies can be associated with several uncertainties - the discreteness of the measurement, the error in the distance to the surface and simulation assumptions regarding the current flow along the wire. From the comparison, we can then calculate the magnetic field induced by the tip at the sample's position. This is important when considering the effect of the magnetic field on spins external to the diamond, such as unpaired electrons on the surface and more broadly a sample placed on the diamond, since by knowing the field generated by the gradient at each point, we can deduce what the Larmor frequency of the target spin would be.

\subsubsection*{Spin imaging resolution}\label{sec: resolution}
To measure the magnetic gradient at higher resolution, we measure ODMR along a line connecting the minima to maxima regions, see Figs.\,\ref{fig:magnetic field}c and \ref{fig:line scan}a. This scan was repeated as a function of distance to the diamond surface, as expected the gradient increases as the tip is closer to the diamond surface, see Fig.\,\ref{fig:line scan}d. Following the law of Biot-Savart (see Supplementary Equation\,\ref{eq: bio-savart}), we fit the increasing gradient to an inverse quadratic function. To obtain the maximal gradient the scan should be performed along the optimal x,y curve, where deviations from this curve will result in a sub-optimal gradient as seen in the point closest to the surface (Fig.\,\ref{fig:line scan}d).
\begin{figure*}[!ht]
    \centering
    \includegraphics[width=\textwidth]{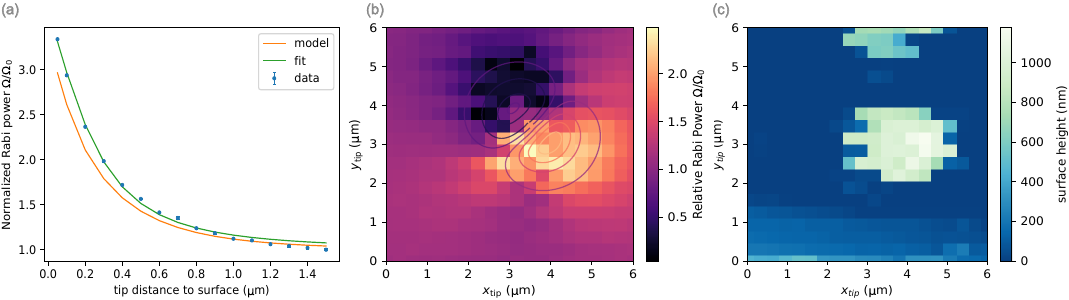}
    \caption{\textbf{Tip-induced Rabi power modulation}. (a) Normalized Rabi power ($\Omega/\Omega_0$) measured for a specific lateral tip position while varying the vertical distance to the surface. Each Rabi measurement is fitted to $Ae^{-\beta t}\cos{(2\pi\nu t)}$ which gives the Rabi period and errors in the plot. Our model (orange curve) shows a similar behavior, with a slight deviation for the closest points. (b) Rabi power as a function of the tip's position relative to the position of the NV, divided by the Rabi power when the tip is far-away from the NV ($\Omega/\Omega_0$), while the distance to the diamond is set to be 156\,nm. (c) A simultaneous AFM scan of the diamond surface where Rabi oscillations were measured. The two raised regions are the nanopillar shape etched in the diamond. The NV used for the measurements is in the lower pillar.}
    \label{fig:rabi}
\end{figure*}
For a current of 1.54\,mA through the tip, the maximum measured gradient is $0.95\,\upmu\text{T nm}^{-1}$, measured when scanning the tip along the axis connecting the minima to maxima regions. The limit on resolution is a combination of the spectral resolution of the setup and the gradient achieved by the tip. With the single NV's dephasing time ($T_2^*$) of $\sim 3\,\upmu\text{s}$, we were able to measure ODMR within an error of 100 kHz. The signal from an electron is detected by the NV at the electron resonance frequency at $\gamma_e B$, where $\gamma_e=28\,\text{MHz mT}^{-1}$ and $B$ is the magnetic field applied to the electron. Thus, to distinguish electrons in our system, the magnetic field difference required is $B=3.6\,\upmu\text{T nm}^{-1}$. While we performed most of the measurements at a relatively low current, a destructive test showed the device can support higher current densities, with a 5-fold improvement in the gradient (see Supplementary Note 5), making it practical for nanometer resolution imaging. In fact, our low-current $\sim 1\,\upmu\text{T nm}^{-1}$ gradient can already distinguish between spin labels, for example, spaced 4.8\,nm in a molecular ruler \cite{Reginsson2012}, with a Larmor precession frequency difference between them of 134\,kHz, as mentioned above, our gradients can be as high as 3.3\,$\upmu\text{T nm}^{-1}$. Additional improvement can be achieved by combining the gradient field with protocols that have demonstrated resolution as good as $\sim\,1\,\text{\AA}$ \cite{Abobeih2019, Yudilevich2022}. Alternatively, the tip itself can be optimized further to improve both geometry and heat load, which may provide a better magnetic gradient.

\subsection*{Tip-induced Rabi power modulation}\label{sec: rabi}
The geometry used here, i.e., where a metallic tip is placed above the diamond while a metallic microwave waveguide is placed below it, is shown to affect the NV Rabi period dramatically. We observe that by a proper positioning of the metallic tip, the Rabi oscillations' frequency ($\Omega$, the Rabi driving power) can be increased for a set MW power applied to the waveguide. Fig.\,\ref{fig:rabi} shows two sets of measurements demonstrating this effect. First, we scan the distance to the diamond surface for a set lateral position and show a relative, $\sim$3.5-fold increase in Rabi power (Fig.\,\ref{fig:rabi}a). Next, Fig.\,\ref{fig:rabi}b shows a lateral scan where the tip is set to a distance of 156\,nm from the surface, and we plot the ratio between the measured Rabi power, $\Omega$, and the Rabi power when the tip is far-away from the NV, $\Omega_0$. Here, there are three noticeable regions - (1) the far-field where the Rabi power is unaffected and equal to the baseline Rabi power as measured without a tip; (2) an enhanced region where the Rabi power increases relative to the baseline, and; (3) a third region where the Rabi power is significantly weaker relative to the baseline. 

The observed modulation of the Rabi power is due to the metallic tip modulating the magnetostatic near-fields that interact with the NV center. Since these are near-field interactions, the modulation's strength is very sensitive to the tip's position. A detailed analytical model, representing the tip geometry using oblate ellipsoids is given in Supplementary Note 9. The contours in Fig.\,\ref{fig:rabi}b are iso-power lines calculated according to the model, and are in agreement with the experimental data. Our model also qualitatively captures the vertical distance dependence (orange line in Fig.\,\ref{fig:rabi}a). When simplifying each ellipsoid to a single point dipole, the data can be fitted by a $z/(z^2+d^2)^{1/2}$ behavior (green curve in Fig.\,\ref{fig:rabi}a), where $z$ is the tip's vertical distance from the NV and $d$ is the lateral one.

The Rabi power modulation is important in several aspects, as working with a larger Rabi frequency provides shorter pulses necessary for more accurate dynamical decoupling sequences \cite{Bar-Gill2013}. Using the tip as a method of increasing the Rabi power, one does not need to ``pay the price'' of thermal management due to increased MW power. When working with the tip, this is particularly valuable since the MW heating can shift the sample when working with strong power. 

These measurements were performed without running current through the tip and are independent of the induced magnetic field. The Rabi power gradient itself, without a current through the tip, can serve as a tool to image spins, since such a large difference in Rabi powers will make it possible to address spins only in a specific region (of the same Rabi power).

\section{Discussion}\label{sec: conclusion}

We introduced a magnetic-field gradient on an AFM tip which can be pulsed with rise/fall times shorter than 600\,ns. This is achieved using a three-step self-aligning deposition process, effectively creating a focused current at the apex of the tip. Using an NV center as a local magnetic sensor, we demonstrated gradients as strong as $1\,\upmu\text{T nm}^{-1}$, which can be even further increased five-fold with a stronger current. Our proof-of-principle of the gradient enables, already now, nanoscale magnetic resonance imaging of proximate unpaired electrons spins in, e.g., a molecular ruler. Moreover, this metallic tip can also serve as a means to locally modulate the driving field around both the target and sensor (NV) spin, yielding an enhancement of $\times 3.5$ in the spin's Rabi driving power. The spatial variability of the Rabi power is and of itself a gradient source for future magnetic resonance imaging in nanometer length scales. The geometry of the tip (apex shape and diameter) as well as choice of deposited metal and the substrate material itself can all be further optimized for an overall higher gradient. Specifically, when working at cryogenic temperatures, a superconducting material can support larger current densities \cite{Molodyk2021} (and no Joule heating), the tip's quartz substrate can be replaced with sapphire or diamond for better heat dissipation \cite{Swoboda2020}, and an optimization of the apex geometry while taking into account Joule heating can yield higher magnetic field gradients in proximity to the tip. 

The current focusing device offers several advantages, overcoming the limitations of existing nanoMRI techniques. First, the ability to pulse the current allows for different magnetic fields during spin manipulation, readout, and initialization, providing optimal NV contrast as defined by the diamond lattice eigenstates. Second, the magnetic field induced by the tip is small relative to the external field and the zero field splitting, meaning only minor shifts are expected, and so the effective Zeeman energy remains largely unchanged when switching between measurements with and without the field from the tip. The possibility to co-align the tip with an arbitrary NV also holds advantages to sensing when working with single NVs. With the geometry presented here, a sample can be placed on the diamond surface in a statistical manner, and once an optimal NV is identified, the gradient can be applied locally around the specific NV, targeting spins and creating a spectral separation between them in Larmor frequency, in the NV's immediate vicinity. Additionally, the current source can also reverse the direction of the magnetic field, enabling the application of gradient pulse sequences similar to those used in conventional NMR \cite{Stejskal1965, Shemesh2013}.

The tip combines two critical features for immediate application in nanoMRI, namely control over the temporal and spatial magnetic field. By working at the upper current limit, we can achieve the gradient necessary for nanoscale imaging. The device thus overcomes the limitations of alternative methods, enabling high-contrast spin read-out with an arbitrary choice of the working NV and of sample placement. Going beyond nanoMRI, we envision our scannable tip with a switchable local magnetic field gradient as a useful tool in active and passive characterization of materials, e.g., (anti)ferromagnets, superconductors and more exotic magnetic phases of matter.

\section*{Methods}\label{sec11}

\subsection*{NV sensor}
The diamond is an e6 [100] electronic-grade CVD diamond, overgrown with a solid state boron rod doping~\cite{Favaro2017}, followed by $^{15}\mathrm{N}^+$ implantation at an energy of 5\,keV and subsequent vacuum annealing at 950$^\circ$C. The boron-doped layer was then etched with oxygen inductively coupled plasma (ICP). The diamond was thinned down to a membrane ($30\,\upmu\mathrm{m}$). Patterns of arrays of nanopillars were etched into the diamond by a combination of electron beam lithography and ICP \cite{Momenzadeh2015}. Optical excitation and detection is performed using a home-built confocal microscope. It comprises a 520\,nm laser for NV initialization and read-out. The diamond sample is scanned with a 2D galvo mirror, which is followed by two lenses. The laser is then focused on the NV with an Olympus MPLFLN100X air-objective with NA 0.9. The fluorescence from the diamond is collected by the objective on to a dichroic mirror and then focused on a 75\,$\upmu\mathrm{m}$ pinhole. The signal is split by a beam splitter to two avalanche photodiodes where the photons are detected, and the counting is facilitated by a Swabian Instruments' TimeTagger20. The positioning of the diamond (x,y,z motion), objective (z motion), permanent magnet (x,y,z motion) and the tip (x,y,z motion) are achieved using an attocube CSFM which provides coarse positioning of each stage. The attocube microscope additionally includes a temperature stability system using a Lakeshore 335 controller. AFM feedback was implemented using frequency-modulation AFM with attocube's ASC500 controller. Tip scanning with respect to the diamond was done using attocube piezo scanners (3$\times$ANSx301).

\subsection*{Tip fabrication}
The process for fabricating the current carrying tip is described in the following. The base material is a 100\,mm-long quartz solid rod (no inner diameter) with two grooves along opposite sides (see Supplementary Note 2). The rod is pulled to form two tips using Sutter Instruments' P-2000, which heats the rod at the center and pulls the two tapers to form two tips. The tips have an apex on the order of 100\,nm and taper length on the order of 5\,mm, where the precise dimensions depend on the working parameters. A home-built thermal deposition \cite{Bagani2019} is used to deposit 2\,nm Cr followed by 10\,nm Au on the two long sides of the rod, creating two leads. To form a 1\,$\upmu\text{m}$ apex, a focused (gallium) ion beam was used to cut the apex by removing the quartz. A second metal deposition was then performed on the two leads and the apex: At the apex, 7 nm\,Cr and 120\,nm Au were deposited and on each of the leads 7\,nm Cr and 50\,nm Au. 

\subsection*{Tip holder and incorporation in the setup}
The tip is placed on a piezo stage which controls its location in x, y and z relative to the diamond sample. An elliptical tip holder was designed to hold the tip at the center with two metal springs that provide electric contact to each lead. Each spring is connected to a coaxial cable to connect to the electric current circuit. A quartz tuning fork ($f_0 = 32.768\,\mathrm{kHz}$, HM International BVBA, model TB-38-20-12.5-32.768) is pressed against the tip using a micropositioner and a live camera, ensuring tip to tuning fork contact and preventing the tuning fork from damaging the tip. 

\subsection*{Co-localization of the tip-NV microscopes}
In order to locate the tip and position it relative to a single NV, a series of scans are performed. First, the diamond sample is removed, and the tip is scanned over the objective, the reflection of the laser from the tip and tuning fork is detected. From this image the tuning fork is identified, and the tip location is estimated within 100 $\upmu\text{m}$ given the known position of the tip relative to the tuning fork. This is followed by a confocal scan of a cross-section of the location of the estimated tip. A second cross-section of the tip close to the apex is scanned to improve the estimation. The tip is then retracted, and the diamond sample is placed between the objective and the tip. Using the AFM feedback, the tip is brought into contact with the diamond sample and retracted up to 5 $\upmu\mathrm{m}$ from the surface. Finally, confocal scans of the diamond membrane show the location of the tip relative to the single NVs. The tip is identified by scanning two images with a small motion of the tip, either in the $x$-$y$ plane or by bringing the tip closer to the sample. For this purpose, working with a tip with an apex larger than the diffraction limit was beneficial.

\subsection*{RF and electrical setup}
The diamond is placed on a co-planar waveguide, the microwave signal is generated by an R\&S SGT100A and modulated up to 400 MHz using IQ mixing signal generated by Spectrum Instrumentation arbitrary waveform generator (AWG) DN2.663-04. The signal is amplified with a Minicircuits ZHL-16W-43-S+ amplifier before injection to the waveguide. The APD signal is collected on the TimeTagger20 (Swabian instruments) which is triggered by the AWG. The signal to the tip is an analog signal generated by the AWG which is amplified using an SRS560 voltage amplifier. 

\subsection*{Data Availability}
The data relevant to figures in the main text are available via the Weizmann/Elsevier data repository, DOI: \href{https://doi.org/10.34933/f3595abc-d114-4ee3-9be1-0739d7d1360d}{10.34933/f3595abc-d114-4ee3-9be1-0739d7d1360d} \cite{ScheinLubomirsky2025data}. Additional raw data are available from the corresponding author upon reasonable request. 

\subsection*{Code Availability}
The MATLAB code is available from the authors upon reasonable request from the corresponding author.

\subsection*{Acknowledgments}
We are thankful to D.\,Yudilevich for providing us with an in-depth proof of the manuscript. We thank E.\,Zeldov and Y.\,Myasoedov for giving us access to their tip puller and thermal deposition tool. We are grateful to U.\,Goldblatt, F.\,Lafont and S.\,Rosenblum for assistance with high-frequency microwave simulations (HFSS). A.\,F.\, is the incumbent of the Elaine Blond Career Development Chair in Perpetuity. We acknowledge support by the Kimmel Institute for Magnetic Resonance. This research is made possible in part by the historic generosity of the Harold Perlman Family. This research was funded exclusively from internal Weizmann Institute of Science resources.

\subsection*{Competing interests}
The authors declare no competing interests.

\subsection*{Author contribution}
LSL built the experimental apparatus, fabricated the tips, ran MATLAB and COMSOL simulations and performed all the measurements and data analysis. YM modeled the tip's effect on the Rabi power. RS and AD implanted the diamond membrane, annealed it and etched nanopillars structures in it. AF conceived and supervised the project. LSL and AF wrote the manuscript with input from YM, RS and AD.

\newpage
\bibliography{main}

\begin{thebibliography}{10}
\expandafter\ifx\csname url\endcsname\relax
  \def\url#1{\burl{#1}}\fi
\expandafter\ifx\csname urlprefix\endcsname\relax\def\urlprefix{URL }\fi
\providecommand{\bibinfo}[2]{#2}
\providecommand{\eprint}[2][]{\url{#2}}
\providecommand{\doi}[1]{\url{https://doi.org/#1}}
\bibcommenthead

\bibitem{Budakian2023}
\bibinfo{author}{Budakian, R.} \emph{et~al.}
\newblock \bibinfo{title}{Roadmap on nanoscale magnetic resonance imaging}.
\newblock \emph{\bibinfo{journal}{Nanotechnology}}
  \textbf{\bibinfo{volume}{35}}, \bibinfo{pages}{412001}
  (\bibinfo{year}{2024}).

\bibitem{suter1992sensitivity}
\bibinfo{author}{Suter, D.}
\newblock \bibinfo{title}{Sensitivity of optically excited and detected
  magnetic resonance}.
\newblock \emph{\bibinfo{journal}{J. Magn. Reson.}}
  \textbf{\bibinfo{volume}{99}}, \bibinfo{pages}{495--506}
  (\bibinfo{year}{1992}).

\bibitem{Ciobanu2002}
\bibinfo{author}{Ciobanu, L.}, \bibinfo{author}{Seeber, D.} \&
  \bibinfo{author}{Pennington, C.}
\newblock \bibinfo{title}{3{D} {MR} microscopy with resolution 3.7
  $\upmu\text{m}$ by 3.3 $\upmu\text{m}$ by 3.3 $\upmu\text{m}$}.
\newblock \emph{\bibinfo{journal}{J. Magn. Reson.}}
  \textbf{\bibinfo{volume}{158}}, \bibinfo{pages}{178--182}
  (\bibinfo{year}{2002}).

\bibitem{Nichol2013}
\bibinfo{author}{Nichol, J.~M.}, \bibinfo{author}{Naibert, T.~R.},
  \bibinfo{author}{Hemesath, E.~R.}, \bibinfo{author}{Lauhon, L.~J.} \&
  \bibinfo{author}{Budakian, R.}
\newblock \bibinfo{title}{Nanoscale {F}ourier-transform magnetic resonance
  imaging}.
\newblock \emph{\bibinfo{journal}{Phys. Rev. X}} \textbf{\bibinfo{volume}{3}},
  \bibinfo{pages}{031016} (\bibinfo{year}{2013}).

\bibitem{Degen2009}
\bibinfo{author}{Degen, C.~L.}, \bibinfo{author}{Poggio, M.},
  \bibinfo{author}{Mamin, H.~J.}, \bibinfo{author}{Rettner, C.~T.} \&
  \bibinfo{author}{Rugar, D.}
\newblock \bibinfo{title}{Nanoscale magnetic resonance imaging}.
\newblock \emph{\bibinfo{journal}{Proc. Natl. Acad. Sci.}}
  \textbf{\bibinfo{volume}{106}}, \bibinfo{pages}{1313--1317}
  (\bibinfo{year}{2009}).

\bibitem{Rose2018}
\bibinfo{author}{Rose, W.} \emph{et~al.}
\newblock \bibinfo{title}{High-resolution nanoscale solid-state nuclear
  magnetic resonance spectroscopy}.
\newblock \emph{\bibinfo{journal}{Phys. Rev. X}} \textbf{\bibinfo{volume}{8}},
  \bibinfo{pages}{011030} (\bibinfo{year}{2018}).

\bibitem{Baumann2015}
\bibinfo{author}{Baumann, S.} \emph{et~al.}
\newblock \bibinfo{title}{Electron paramagnetic resonance of individual atoms
  on a surface}.
\newblock \emph{\bibinfo{journal}{Science}} \textbf{\bibinfo{volume}{350}},
  \bibinfo{pages}{417--420} (\bibinfo{year}{2015}).

\bibitem{Sellies_2023}
\bibinfo{author}{Sellies, L.} \emph{et~al.}
\newblock \bibinfo{title}{Single-molecule electron spin resonance by means of
  atomic force microscopy}.
\newblock \emph{\bibinfo{journal}{Nature}} \textbf{\bibinfo{volume}{624}},
  \bibinfo{pages}{64--68} (\bibinfo{year}{2023}).

\bibitem{Gruber1997}
\bibinfo{author}{Gruber, A.} \emph{et~al.}
\newblock \bibinfo{title}{Scanning confocal optical microscopy and magnetic
  resonance on single defect centers}.
\newblock \emph{\bibinfo{journal}{Science}} \textbf{\bibinfo{volume}{276}},
  \bibinfo{pages}{2012--2014} (\bibinfo{year}{1997}).

\bibitem{Balasubramanian2009}
\bibinfo{author}{Balasubramanian, G.} \emph{et~al.}
\newblock \bibinfo{title}{Ultralong spin coherence time in isotopically
  engineered diamond}.
\newblock \emph{\bibinfo{journal}{Nat. Mater.}} \textbf{\bibinfo{volume}{8}},
  \bibinfo{pages}{383--387} (\bibinfo{year}{2009}).

\bibitem{Grinolds2013}
\bibinfo{author}{Grinolds, M.~S.} \emph{et~al.}
\newblock \bibinfo{title}{Nanoscale magnetic imaging of a single electron spin
  under ambient conditions}.
\newblock \emph{\bibinfo{journal}{Nat. Phys.}} \textbf{\bibinfo{volume}{9}},
  \bibinfo{pages}{215--219} (\bibinfo{year}{2013}).

\bibitem{Staudacher2013}
\bibinfo{author}{Staudacher, T.} \emph{et~al.}
\newblock \bibinfo{title}{Nuclear magnetic resonance spectroscopy on a
  (5-nanometer)$^3$ sample volume}.
\newblock \emph{\bibinfo{journal}{Science}} \textbf{\bibinfo{volume}{339}},
  \bibinfo{pages}{561--563} (\bibinfo{year}{2013}).

\bibitem{Mamin2013}
\bibinfo{author}{Mamin, H.~J.} \emph{et~al.}
\newblock \bibinfo{title}{Nanoscale nuclear magnetic resonance with a
  nitrogen-vacancy spin sensor}.
\newblock \emph{\bibinfo{journal}{Science}} \textbf{\bibinfo{volume}{339}},
  \bibinfo{pages}{557--560} (\bibinfo{year}{2013}).

\bibitem{Sushkov2014}
\bibinfo{author}{Sushkov, A.~O.} \emph{et~al.}
\newblock \bibinfo{title}{Magnetic resonance detection of individual proton
  spins using quantum reporters}.
\newblock \emph{\bibinfo{journal}{Phys. Rev. Lett.}}
  \textbf{\bibinfo{volume}{113}}, \bibinfo{pages}{197601}
  (\bibinfo{year}{2014}).

\bibitem{Mueller2014}
\bibinfo{author}{M\"{u}ller, C.} \emph{et~al.}
\newblock \bibinfo{title}{Nuclear magnetic resonance spectroscopy with single
  spin sensitivity}.
\newblock \emph{\bibinfo{journal}{Nat. Commun.}} \textbf{\bibinfo{volume}{5}},
  \bibinfo{pages}{4703} (\bibinfo{year}{2014}).

\bibitem{Taminiau2012}
\bibinfo{author}{Taminiau, T.~H.} \emph{et~al.}
\newblock \bibinfo{title}{Detection and control of individual nuclear spins
  using a weakly coupled electron spin}.
\newblock \emph{\bibinfo{journal}{Phys. Rev. Lett.}}
  \textbf{\bibinfo{volume}{109}}, \bibinfo{pages}{137602}
  (\bibinfo{year}{2012}).

\bibitem{Kolkowitz2012}
\bibinfo{author}{Kolkowitz, S.}, \bibinfo{author}{Unterreithmeier, Q.~P.},
  \bibinfo{author}{Bennett, S.~D.} \& \bibinfo{author}{Lukin, M.~D.}
\newblock \bibinfo{title}{Sensing distant nuclear spins with a single electron
  spin}.
\newblock \emph{\bibinfo{journal}{Phys. Rev. Lett.}}
  \textbf{\bibinfo{volume}{109}}, \bibinfo{pages}{137601}
  (\bibinfo{year}{2012}).

\bibitem{Zopes2018}
\bibinfo{author}{Zopes, J.} \emph{et~al.}
\newblock \bibinfo{title}{Three-dimensional localization spectroscopy of
  individual nuclear spins with sub-angstrom resolution}.
\newblock \emph{\bibinfo{journal}{Nat. Commun.}} \textbf{\bibinfo{volume}{9}},
  \bibinfo{pages}{4678} (\bibinfo{year}{2018}).

\bibitem{Grinolds2014}
\bibinfo{author}{Grinolds, M.~S.} \emph{et~al.}
\newblock \bibinfo{title}{Subnanometre resolution in three-dimensional magnetic
  resonance imaging of individual dark spins}.
\newblock \emph{\bibinfo{journal}{Nat. Nanotechnol.}}
  \textbf{\bibinfo{volume}{9}}, \bibinfo{pages}{279--284}
  (\bibinfo{year}{2014}).

\bibitem{Balasubramanian2008}
\bibinfo{author}{Balasubramanian, G.} \emph{et~al.}
\newblock \bibinfo{title}{Nanoscale imaging magnetometry with diamond spins
  under ambient conditions}.
\newblock \emph{\bibinfo{journal}{Nature}} \textbf{\bibinfo{volume}{455}},
  \bibinfo{pages}{648--651} (\bibinfo{year}{2008}).

\bibitem{Tetienne2012}
\bibinfo{author}{Tetienne, J.-P.} \emph{et~al.}
\newblock \bibinfo{title}{Magnetic-field-dependent photodynamics of single nv
  defects in diamond: an application to qualitative all-optical magnetic
  imaging}.
\newblock \emph{\bibinfo{journal}{New J. Phys.}} \textbf{\bibinfo{volume}{14}},
  \bibinfo{pages}{103033} (\bibinfo{year}{2012}).

\bibitem{Bodenstedt2018}
\bibinfo{author}{Bodenstedt, S.} \emph{et~al.}
\newblock \bibinfo{title}{Nanoscale spin manipulation with pulsed magnetic
  gradient fields from a hard disc drive writer}.
\newblock \emph{\bibinfo{journal}{Nano Lett.}} \textbf{\bibinfo{volume}{18}},
  \bibinfo{pages}{5389--5395} (\bibinfo{year}{2018}).

\bibitem{Arai2015}
\bibinfo{author}{Arai, K.} \emph{et~al.}
\newblock \bibinfo{title}{Fourier magnetic imaging with nanoscale resolution
  and compressed sensing speed-up using electronic spins in diamond}.
\newblock \emph{\bibinfo{journal}{Nat. Nanotechnol.}}
  \textbf{\bibinfo{volume}{10}}, \bibinfo{pages}{859--864}
  (\bibinfo{year}{2015}).

\bibitem{Amawi2023}
\bibinfo{author}{Amawi, M.~T.} \emph{et~al.}
\newblock \bibinfo{title}{Three-dimensional magnetic resonance tomography with
  sub-10 nanometer resolution}.
\newblock \emph{\bibinfo{journal}{npj Quantum Inf.}}
  \textbf{\bibinfo{volume}{10}}, \bibinfo{pages}{16} (\bibinfo{year}{2024}).

\bibitem{Wang2023}
\bibinfo{author}{Guo, Z.} \emph{et~al.}
\newblock \bibinfo{title}{Wide-field fourier magnetic imaging with electron
  spins in diamond}.
\newblock \emph{\bibinfo{journal}{npj Quantum Inf.}}
  \textbf{\bibinfo{volume}{10}}, \bibinfo{pages}{24} (\bibinfo{year}{2024}).

\bibitem{Stejskal1965}
\bibinfo{author}{Stejskal, E.~O.} \& \bibinfo{author}{Tanner, J.~E.}
\newblock \bibinfo{title}{Spin diffusion measurements: Spin echoes in the
  presence of a time-dependent field gradient}.
\newblock \emph{\bibinfo{journal}{J. Chem. Phys.}}
  \textbf{\bibinfo{volume}{42}}, \bibinfo{pages}{288--292}
  (\bibinfo{year}{1965}).

\bibitem{Shemesh2013}
\bibinfo{author}{Shemesh, N.}, \bibinfo{author}{Álvarez, G.~A.} \&
  \bibinfo{author}{Frydman, L.}
\newblock \bibinfo{title}{Measuring small compartment dimensions by probing
  diffusion dynamics via non-uniform oscillating-gradient spin-echo ({NOGSE})
  {NMR}}.
\newblock \emph{\bibinfo{journal}{J. Magn. Reson.}}
  \textbf{\bibinfo{volume}{237}}, \bibinfo{pages}{49--62}
  (\bibinfo{year}{2013}).

\bibitem{Edmonds2021}
\bibinfo{author}{Edmonds, A.~M.} \emph{et~al.}
\newblock \bibinfo{title}{Characterisation of {CVD} diamond with high
  concentrations of nitrogen for magnetic-field sensing applications}.
\newblock \emph{\bibinfo{journal}{Mater. Quantum Technol.}}
  \textbf{\bibinfo{volume}{1}}, \bibinfo{pages}{025001} (\bibinfo{year}{2021}).

\bibitem{Hong2012}
\bibinfo{author}{Hong, S.} \emph{et~al.}
\newblock \bibinfo{title}{Coherent, mechanical control of a single electronic
  spin}.
\newblock \emph{\bibinfo{journal}{Nano Lett.}} \textbf{\bibinfo{volume}{12}},
  \bibinfo{pages}{3920} (\bibinfo{year}{2012}).

\bibitem{Reginsson2012}
\bibinfo{author}{Reginsson, G.~W.}, \bibinfo{author}{Kunjir, N.~C.},
  \bibinfo{author}{Sigurdsson, S.~T.} \& \bibinfo{author}{Schiemann, O.}
\newblock \bibinfo{title}{Trityl radicals: Spin labels for nanometer-distance
  measurements}.
\newblock \emph{\bibinfo{journal}{Chem. Eur. J.}}
  \textbf{\bibinfo{volume}{18}}, \bibinfo{pages}{13580--13584}
  (\bibinfo{year}{2012}).

\bibitem{Abobeih2019}
\bibinfo{author}{Abobeih, M.~H.} \emph{et~al.}
\newblock \bibinfo{title}{Atomic-scale imaging of a 27-nuclear-spin cluster
  using a single-spin quantum sensor}.
\newblock \emph{\bibinfo{journal}{Nature}} \textbf{\bibinfo{volume}{576}},
  \bibinfo{pages}{411--415} (\bibinfo{year}{2019}).

\bibitem{Yudilevich2022}
\bibinfo{author}{Yudilevich, D.}, \bibinfo{author}{St\"ohr, R.},
  \bibinfo{author}{Denisenko, A.} \& \bibinfo{author}{Finkler, A.}
\newblock \bibinfo{title}{Mapping single electron spins with magnetic
  tomography}.
\newblock \emph{\bibinfo{journal}{Phys. Rev. Applied}}
  \textbf{\bibinfo{volume}{18}}, \bibinfo{pages}{054016}
  (\bibinfo{year}{2022}).

\bibitem{Bar-Gill2013}
\bibinfo{author}{Bar-Gill, N.}, \bibinfo{author}{Pham, L.},
  \bibinfo{author}{Jarmola, A.}, \bibinfo{author}{Budker, D.} \&
  \bibinfo{author}{Walsworth, R.}
\newblock \bibinfo{title}{Solid-state electronic spin coherence time
  approaching one second}.
\newblock \emph{\bibinfo{journal}{Nat. Commun.}} \textbf{\bibinfo{volume}{4}},
  \bibinfo{pages}{1743} (\bibinfo{year}{2013}).

\bibitem{Molodyk2021}
\bibinfo{author}{Molodyk, A.} \emph{et~al.}
\newblock \bibinfo{title}{Development and large volume production of extremely
  high current density {YB}a$_{2}${C}u$_{3}${O}$_{7}$ superconducting wires for
  fusion}.
\newblock \emph{\bibinfo{journal}{Sci. Rep.}} \textbf{\bibinfo{volume}{11}},
  \bibinfo{pages}{2084} (\bibinfo{year}{2021}).

\bibitem{Swoboda2020}
\bibinfo{author}{Swoboda, T.}, \bibinfo{author}{Klinar, K.},
  \bibinfo{author}{Yalamarthy, A.~S.}, \bibinfo{author}{Kitanovski, A.} \&
  \bibinfo{author}{Muñoz~Rojo, M.}
\newblock \bibinfo{title}{Solid‐state thermal control devices}.
\newblock \emph{\bibinfo{journal}{Adv. Electron. Mater.}}
  \textbf{\bibinfo{volume}{7}}, \bibinfo{pages}{202000625}
  (\bibinfo{year}{2021}).

\bibitem{Favaro2017}
\bibinfo{author}{F\'avaro~de Oliveira, F.} \emph{et~al.}
\newblock \bibinfo{title}{Tailoring spin defects in diamond by lattice
  charging}.
\newblock \emph{\bibinfo{journal}{Nat. Commun.}} \textbf{\bibinfo{volume}{8}},
  \bibinfo{pages}{15409} (\bibinfo{year}{2017}).

\bibitem{Momenzadeh2015}
\bibinfo{author}{Momenzadeh, S.~A.} \emph{et~al.}
\newblock \bibinfo{title}{Nanoengineered diamond waveguide as a robust bright
  platform for nanomagnetometry using shallow nitrogen vacancy centers}.
\newblock \emph{\bibinfo{journal}{Nano Lett.}} \textbf{\bibinfo{volume}{15}},
  \bibinfo{pages}{165--169} (\bibinfo{year}{2014}).

\bibitem{Bagani2019}
\bibinfo{author}{Bagani, K.} \emph{et~al.}
\newblock \bibinfo{title}{Sputtered {M}o$_{66}${R}e$_{34}$ {SQUID}-on-tip for
  high-field magnetic and thermal nanoimaging}.
\newblock \emph{\bibinfo{journal}{Phys. Rev. Applied}}
  \textbf{\bibinfo{volume}{12}}, \bibinfo{pages}{044062}
  (\bibinfo{year}{2019}).

\bibitem{ScheinLubomirsky2025data}
\bibinfo{author}{Schein-Lubomirsky, L.}
\newblock \bibinfo{title}{Data for: Pulsed magnetic field gradient on a tip for
  nanoscale imaging of spins}.

\end{thebibliography}

\end{document}